\documentclass{emulateapj}
\usepackage{natbib,graphicx,psfig}

\def\refnew#1{(\ref{#1})}

\def \yrs {\, \rm yrs}

\def \m {\, \rm m}

\def \s {\, \rm s}

\def \be {\begin{equation}}
\def \ee {\end{equation}}

\def\lesssim{\mathrel{\hbox{\rlap{\hbox{\lower4pt\hbox{$\sim$}}}\hbox{$<$}}}}
\def\gtrsim{\mathrel{\hbox{\rlap{\hbox{\lower4pt\hbox{$\sim$}}}\hbox{$>$}}}}
\newcommand       \bea          {\begin{eqnarray}}
\newcommand       \eea          {\end{eqnarray}}

\newcommand       \au           {\rm \, AU}

\begin{document}

\title{Hot Jupiters in binary star systems}

\author{Yanqin Wu\altaffilmark{1}, Norman W. Murray\altaffilmark{2,3} 
\& J. Michael Ramsahai\altaffilmark{2}}
\altaffiltext{1}{Department of Astronomy \& Astrophysics, 
 University of Toronto, Toronto, ON M5S 3H4, Canada}
\altaffiltext{2}{Canadian Institute of Theoretical Astrophysics,
 University of Toronto, Toronto, ON M5S 3H8, Canada}
\altaffiltext{3}{Canada Research Chair in Astrophysics}
\email{wu@astro.utoronto.ca; murray@cita.utoronto.ca}

\begin{abstract}

Radial velocity surveys find Jupiter mass planets with semi-major axes
$a$ less than 0.1 AU around $\sim 1\%$ of solar-type stars;  %
counting planets with $a$ as large as 5 AU, the fraction of stars
having planets reaches $\sim 10\%$ \citep{Marcy,Butler}. An  %
examination of the distribution of semi-major axes shows that there is
a clear excess of planets with orbital periods around 3 or 4 days,
corresponding to $a\approx 0.03$ AU, with a sharp cutoff at shorter
periods (see Figure 1). It is believed that Jupiter mass planets form
at large distances from their parent stars; some fraction then migrate
in to produce the short period objects.  We argue that a significant
fraction of the `hot Jupiters' ($ a<0.1$AU) may arise in binary star
systems in which the orbit of the binary is highly inclined to the
orbit of the planet. Mutual torques between the two orbits drive down
the minimum separation or periapse $r_p$ between the planet and its
host star (the Kozai mechanism). This periapse collapse is halted when
tidal friction on the planet circularizes the orbit faster than Kozai
torque can excite it. The same friction then circularizes the planet
orbit, producing hot Jupiters with the peak of the semimajor axis
distribution lying around 3 days. For the observed distributions of
binary separation, eccentricity and mass ratio, roughly $ 2.5\%$ of
planets with initial semimajor axis $a_p\approx 5\au$ will migrate to
within $0.1\au$ of their parent star. Kozai migration could
account for $10\%$ or more of the observed hot Jupiters. %
\end{abstract}


\section{Introduction to Kozai Migration}
\label{sec:intro}

Statistics from radial velocity planet searches \citep{Marcy,Butler}
show that
the occurrence rate of giant planets within 0.1 AU (``hot-Jupiters'')
is $\sim 1\%$; extrapolating to 20 AU the occurrence is $12\%$.  There
is a clear "pile-up" of planets with orbital periods near 3 days
(Fig. \ref{fig:fig1}). Transit observations yield a similar fraction
of hot Juptiers \citep{Gould,Fressin}. %
What migration mechanisms can produce the observed feature in semi-major %
axis distributions represented by hot Jupiters? In this article we
focus on the mechanism known as Kozai migration. %

Consider a planet circling a star that is a member of a binary %
system. The mutual torques between the binary and planetary orbits
transfer angular momentum between the two while leaving the orbital
energies nearly unchanged. For mutual inclinations $I\gtrsim 40^\circ$
a resonance between the precession rate of the planet's nodal and
apsidal lines greatly enhances the effectiveness of this exchange of
angular momentum, producing large oscillations in the planet's angular
momentum \citep[Kozai cycles,][]{Kozai}.  The planet eccentricity %
($e_p$) and periapse ($r_p\equiv a_p(1-e_p)$) oscillate with a
characteristic timescale \citep{HolmanTouma} 
\be 
P_{\rm Kozai} \approx
{m_*\over{m_c}} {{P_c^2}\over{P_p}} (1-e_c^3)^{3/2},
\label{eq:kozaip}
\ee  
where $m_*$ and $m_c$ are the masses of the central and companion
stars, while $P_c$ and $P_p$ are the periods of the binary and
planetary orbits, respectively.  The binary eccentricity is denoted by
$e_c$.
\citet{HolmanTouma} and \citet{Takeda}, among others, have studied the %
role of these Kozai cycles in producing the eccentricities observed in
known exo-planets.

For sufficiently large $I$, $r_p$ can reach very small values,
allowing tidal dissipation to erode the orbit of the planet. 
\citet{EKE} were the first to propose that Kozai cycles, in
combination with tidal friction, can shrink the orbit of a inner
binary in a hierarchical triple system, leading to the formation of
contact binaries. \citet[][hereafter WM03]{WuMurray} have studied 
Kozai migration in application to exo-planets and found it to be the
only plausible explanation for the migration of the planetary object
HD80606b.

In the absence of any other modification of the gravitational
potential, the minimum $r_p$ may fall below the Roche radius ($r_R$)
and the planet may be destroyed.  However, there are a number of
competing torques that can limit the amount of angular momentum that
the Kozai torque can extract from the orbit of the planet, including
general relativistic (GR) corrections to Newtonian gravity, and
torques associated with the extended mass distribution of both the
primary star and the planet. The latter includes rotationally induced
planetary oblateness, the tidal bulge raised by the star on the
planet, the misalignment of this bulge produced by friction, and the
stellar counterparts of all these. These torques can halt the
Kozai-induced collapse in $r_p$ and promote planetary survival.

Which torque becomes competitive with the Kozai torque depends on the
system; for systems with very large binary semi-major axis ($a_c$) and
therefore very weak Kozai torque, the GR precession can halt the
reduction of $r_p$ before tides become important. However, for tighter
or more inclined binaries,
tidal friction sets the minimum $r_p$. Since the tidal torques depend
strongly on $r_p$, binary systems with a wide range of $a_c/a_p$ will
be stalled at essentially the same $r_p$, leading to a pile-up of hot
Jupiters at $a_p \sim 2 r_p$ when the planet orbits are later
circularized.

\section{Numerical Experiments}
\label{sec:experiment}

We quantify the effect of Kozai migration by considering an ensemble
of binary systems following that in
\citet{Takeda}.  These binaries are initially comprised of a solar-mass host star, a
jupiter-mass planet ($m_p = M_J$) orbiting at $5$ AU with an
eccentricity of $0.05$, and a binary companion of mass $0.23 M_\odot$
-- this is the peak of the observed mass ratio distribution in the
solar neighbourhood
\citep{Mayor}. The distribution in binary separation ($P(a_c)$) is assumed flat in
logarithmic $a_c$ ($a_c$ ranging from $20$ to $20,000$ AU). We set
$P(e_b) = 2 e_b$, a thermal distribution often adopted in binary
population synthesis. This latter choice hardly affects the
results. The last ansatz, our most sensitive yet most uncertain
assumption, takes $I$ to be isotropically distributed. Based on
studies of stellar spin and binary orbits
\citep{Hale}, this seems reasonable for $a_c>40$ AU, but may be less
appropriate for tighter binaries; polarimetry studies of protostellar
disks suggest that the circumstellar disk and the binary plane are
correlated for $a_c$ up to a few hundred AU
\citep{Jensen,Monin}. However, polarimetry estimates only
the projected angle between the two planes, and is strongly plagued by
interstellar polarization. The results should be taken with caution at
present.

\begin{figure}[t]
\centerline{\psfig{figure=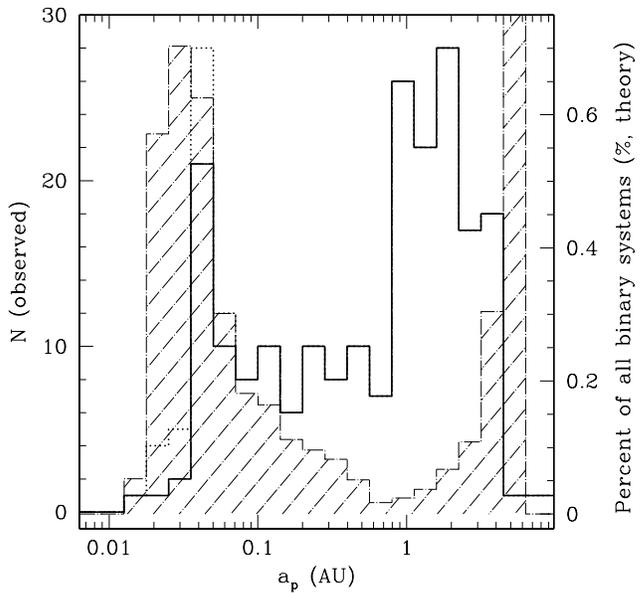,width=1.00\hsize}}
\caption[]{The histogram of the planet semi-major axis (logarithmic)
distribution. The thick solid curve is the observed radial velocity
planet distribution.  The planets detected by the transit technique
are added on top (dotted line) assuming that the detection
efficiencies are the same between the two techniques. The shaded area
shows the simulation result, with the vertical axis read at the
right. The peak at $\sim 3$ day orbital period corresponds to planets
that are Kozai migrated and later circularized.  The position and
width of this peak depends on a number of parameters (see
Eq. [\ref{eq:bcrit}]).  In particular, if planet radius is a
decreasing function of planet age, the width of the peak shrinks (see
Fig. \ref{fig:fig2}).  The narrow peak at $\sim 5$ AU corresponds to
planets that are unmigrated, remaining at their initial $a_p$.}
\label{fig:fig1}
\end{figure}

We produce an ensemble of $100,000$ systems. Out of these we %
select systems that can potentially perturb the planet to $r_p\lesssim
0.1$ AU. To reach this distance, a planet starting at semi-major axis
$a_p$ (with a small eccentricity) will have to attain $e_{\rm max}
\geq 1-0.1/a_p$. Ignoring tidal dissipation,\footnote{Tidal
dissipation increases the Kozai integral and slightly raises the
minimum requirement on $I$ (WM03).} the Kozai integral (the planet's
orbital angular momentum in the normal of the binary plane) $H_K =
\sqrt{(1-e_p^2)} \cos I = {\rm constant}$. Taking a minimum $I
\approx 40 \deg$ during the Kozai cycles \citep[see,
e.g.,][]{HolmanTouma}, this yields a minimum initial inclination
required for producing hot Jupiters: $I \gtrsim 81 \deg$. This value
is independent of the binary separation or mass.  The fraction of
isotropically inclined systems that have such a misalignment is $\sim
15\%$.

We then weed out planets that are likely dynamically unstable
according to the following fitting formula 
\be
{{a_p}\over{a_c}} \geq 0.330-0.417 e_c+0.069 e_c^2.
\label{eq:apmax-vert}
\ee 
This expression is obtained by integrating the orbits of our initial
system for $10^4$ binary orbits, taking $I = 90 \deg$.  This
non-coplanar stability limit is $15\%$ to $30\%$ more restrictive than
the coplanar stability limit found by \citet{HolmanWiegert}. It is
used here as a rough proxy for systems that either eject their planets
quickly after formation, or are unable to form planets due to the
strong tidal influence of the companion.  This proceedure eliminates
many systems with $a_c < 100$ AU; we are left with $\sim 10\%$ of the
original ensemble that could potentially reach $< 0.1$ AU, if they are
not stalled by other torques at larger distances.

These remaining systems are integrated using secular equations
obtained by averaging over the orbital motions of both the planet and
the binary companion \citep{EKE}. These equations include the effects
of Kozai perturbation, tidal dissipation, GR precession, and tidal and
rotational bulge precessions.\footnote{In this study, we rely
exclusively on these secular equations. The actual dynamics may
deviate due to short term noises and mean-motion perturbations and
should be studied with N-body integration codes.} We use a Runge-Kutta
integrator with an adaptive step size set to keep the integration
error below a preset limit. We follow the procedure described in
WM03, which also lists values for the various parameters involved. In
particular, we choose the initial stellar spin direction to be aligned
with the initial orbit normal for the planet.

The integration is stopped after $5$ Gyrs have passed, or when $5$
million timesteps are exhausted, or when $r_p<2 R_\odot$. The last
condition roughly corresponds to the planet overflowing its Roche %
lobe; however, {\it none} of the planet in our simulation reached this
state.\footnote{This is due to the strong dependence of the tidal timescale
on $r_p$; tidal distortions act as a barrier,
maintaining $r_p\gtrsim r_R$.}  %
The limit on the number of integration timesteps is usually reached if
Kozai oscillations have been effectively halted by rapid tidal or
other precessions; in that case the subsequent dynamical evolution of
the planet simply reduces $e_p$.  We then use a simplified code,
including only the effects of tidal dissipation on the planet orbit
and planet/stellar spins, to finish integrating to $5$ Gyrs.

We find that about $2.5\%$ of our ensemble eventually migrate %
inward of $0.1$ AU. The distribution of final semi-major axes is
concentrated between $0.02$ AU and $0.05$ AU with a peak at $0.03$
AU. Our hot Jupiters exhibit a pile-up at $\sim3$ day periods similar
to the observed population (Fig. \ref{fig:fig1}). %

Given the same initial $I$, tighter binaries produce a closer-in hot Jupiter
in a shorter amount of time.  Many of the hot Jupiters are tidally
ensnared on their first close approach to the host star
(Fig. \ref{fig:fig3}), with the Kozai period between $10^4$ to $10^8$
yrs.  Tidal circularization of these orbits then takes upward of
$10^7$ years.

\begin{figure}[t]
\centerline{\psfig{figure=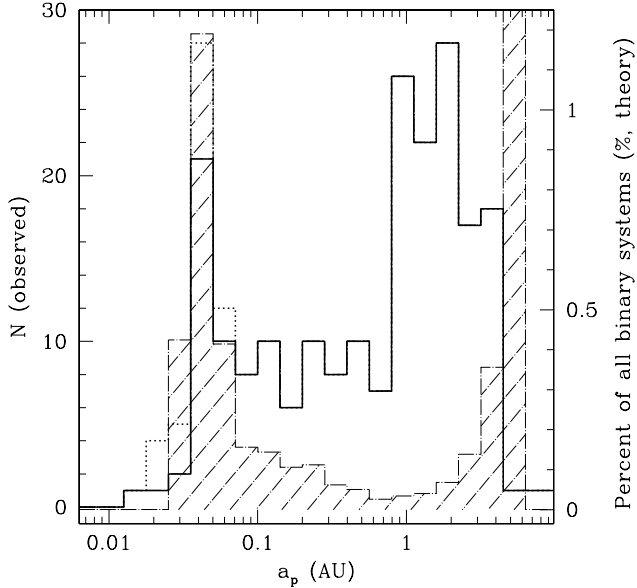,width=1.00\hsize}}
\caption[]{Similar to Fig. \ref{fig:fig1}, except where we have taken the 
planet radius to shrink as $R_p = R_J [1 + \exp(-t/3\times 10^7 yrs)]$. 
The 3-day feature narrows significantly.}
\label{fig:fig2}
\end{figure}

\begin{figure}[t]
\centerline{\psfig{figure=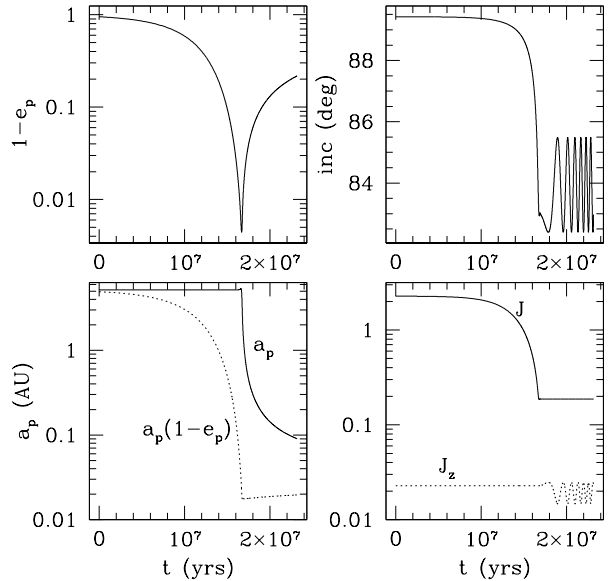,width=1.00\hsize}}
\caption[]{Migration history for a system that in the absence of tides
would have reached a minimum distance of $0.0004$ AU ($0.1 R_\odot$)
and been declared lost; in the presences of tides it
reaches a minimum distance of $0.013$ AU and
is later circularized at $a_p = 0.026$ AU. The four panels %
are: top-left, planet eccentricity as a function of time (in years);
top-right, relative inclination between the two orbit normals;
bottom-left, planet semi-major axis (solid, in AU) and periapse
(dotted); bottom-right, planet total orbital angular momentum ($J$,
solid) and its component along the orbit normal of the binary ($J_z$,
dotted), both in arbitrary units.  Kozai oscillation (which conserves
$J_z$) has proceeded for barely half a cycle before the orbital energy
of the planet is significantly dissipated and the planet is removed
from the influence of the binary companion.  Tidal dissipation
operates afterwards (during which $J$ is conserved). The inclination
angle evolves little in this example. }
\label{fig:fig3}  
\end{figure}

The 3-day feature in the computed $a_p$ distribution appears wider
than the observed distribution. However, as just noted, closer-in
planets are migrated in earlier, so they still have larger radii and
larger stalling peraipses. Experimenting with the following time
evolution of planet radius,
\be
R_p = R_J \left[ 1 + \exp\left(- {t\over{\tau_{\rm shrink}}}\right)\right],
\ee
with $\tau_{\rm shrink}$ taken to be $3\times 10^7 \yrs$, we find that
the 3-day bump narrows significantly (Fig. \ref{fig:fig2}).

\section{Discussion}
\label{sec:discuss}

\subsection{Stalling Radius and the 3-day Pile-up}
\label{subsec:stall}

\begin{figure}[t]
\centerline{\psfig{figure=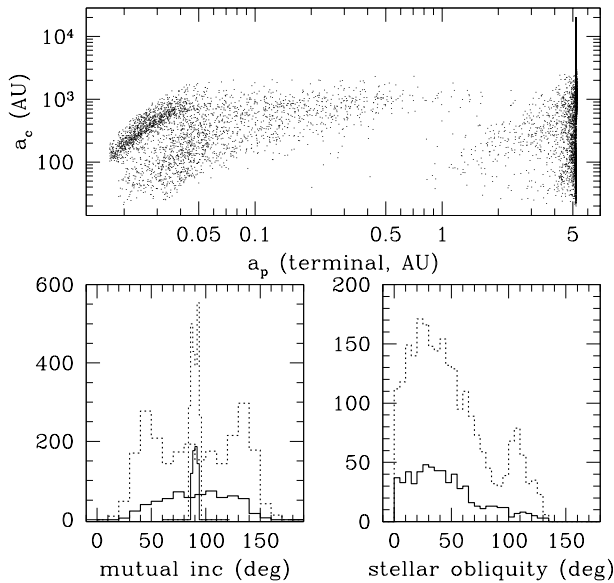,width=1.00\hsize}}
\caption[]{Parameters for the binary systems that produce Kozai
migration. In the top panel, the final $a_p$ (horizontal
axis) is plotted against $a_c$.  Smaller values of the former are in
general correlated with closer binaries (eq. \ref{eq:bcrit}), with
most hot Jupiters arising from binaries with $a_c \in [100,1000]$ AU.
The bottom left panel shows the distributions of initial (thin lines)
and final (thicker lines) inclinations between the two orbital planes
-- the solid curves include systems with $a_p<0.025$ AU, and the
dotted curves all systems with final $a_p< 0.1$ AU, (similarly in the
right panel). The final inclination angles are much more spread out,
as the Kozai cycles convert inclination to eccentricity --
\citet{Fabrycky} gives a detailed explanation for the features. The
bottom right panel shows the distribution of final angles $\psi$
between the stellar spin axis and the planetary orbit normal. Most
systems (especially the tightest ones) have $\psi<50^\circ$, although
some stars may spin retrograde relative to the planet orbit.
}
\label{fig:fig4}
\end{figure}

The periapse of a Kozai-migrating planet is stalled at a distance where the
eccentricity forcing due to the binary companion is counteracted by
the eccentricity damping by tidal dissipation.  Kozai forcing yields
\citep{EKE} 
\be
{1\over e_p} {{de_p}\over{dt}} \approx 5 (1-e_p^2) {{m_c
n_c^2}\over{m_p+m_*+m_c}} {1\over{4 n_p \sqrt{1-e_p^2}
(1-e_c^2)^{3/2}}},
\label{eq:taue}
\ee
where $n_c = 2\pi/P_c$, $n_p = 2\pi/P_p$. The rate of tidal
eccentricity damping depends strongly on the periapse distance. Considering
only tides raised on the planet, we obtain\citep{Hut1981}
\be 
{1\over e_p} {{de_p}\over{dt}} \approx - {{27 k_p G m_p}\over{2 R_p^3
Q_p n_p}} {1\over q}\left(1+{1\over q}\right)
\left({{R_p}\over{a_p}}\right)^8 {1\over{(1-e_p^2)^{13/2}}},
\label{eq:edothut}
\ee 
where $k_p$ is the planet's tidal Love number, $Q_p$ its tidal
dissipation factor and $R_p$ its radius (see WM03). The mass ratio $q
= m_p/m_*$.  Equating the two rates, we obtain the stalling periapse
value,
\begin{eqnarray}
r_{p, \rm stall} & = & 0.015 AU \left[
\left({{m_*}\over{M_\odot}}\right)^{3/2}\,
\left({{M_J}\over{m_p}}\right)\,
\left({{3 \times 10^5}\over{Q_p}}\right)\,
\left({{k_p}\over{0.5}}\right)\, 
\right. \nonumber \\
& & \hskip-0.3in \left. \times  \left({{R_p}\over{R_J}}\right)^5\,
\left({{5 AU}\over{a_p}}\right)\,
\left({{0.23 M_\odot}\over{m_c}}\right)\,
\left({{a_c}\over{270 AU}}\right)^3\,
\right]^{1/6.5},
\label{eq:bcrit}
\end{eqnarray}
where we have scaled variables by their representative values ($R_J$
is the radius of Jupiter). Coincidentally, $r_{p,\rm stall} \sim r_R$,
and it depends little on a variety of parameters, including stellar
mass, companion mass, planet mass, planet tidal $Q$ factor, and planet
initial orbit. This justifies our choices for these parameters in the
numerical experiment.
\footnote{This mechanism works for other types of planets like hot
Neptunes or super-earths. Substituting into Eq. \refnew{eq:bcrit}
values appropriate for Neptune and Earth,
we obtain similar values for $r_{p,\rm stall}$.} %
In our simulation, most binaries that give rise to hot Jupiters have
$a_c \in [100,1000]$ AU (Fig. \ref{fig:fig4}) and we have scaled $a_c$
here by roughly the median value.  Tighter binaries are relatively
unimportant -- planets in many of these systems are dynamically
unstable and are excluded from our study. 

In the subsequent tidal circularization, orbital angular momentum is
roughly conserved and the final $a_p \sim 2 r_{p, \rm stall}
\sim 0.03$ AU.

The fraction of stars with Kozai migrated hot Jupiters is given by
\be  
f_{< 0.1} = f_b\cdot f_p\cdot f_{Kozai},
\ee  
where $f_b$ is the fraction of stars in binary systems, $f_p$ is the
fraction of solar type stars with Jupiter mass planets formed at a few
AU, and $f_{\rm Kozai}$ is the fraction of planets in binary star
systems that undergo Kozai migration to $a_p<0.1$AU.  Taking $f_b
\approx 0.65$ \citep{Mayor},
$f_p \gtrsim 0.07$ \citep{Marcy}, and $f_{\rm Kozai} \approx0.025$
(this work), we suggest that, at a minimum, $10\%$ of the known hot
Jupiters may be due to Kozai migration. The most uncertain number is
$f_p$.
The value of $f_p$ we have quoted is the observed fraction in the Keck
sample, which is substantially complete up to $a_p\approx 3\au$.
Assuming the number of planets per $\au$ is flat up to $a_p=30\au$
gives $f_p=0.12$ and $f_{< 0.1}=0.002$. There is some indication that
the number of planets per $\au$ is an increasing function of $a_p$.
If $f_p=0.5$, more than half the hot Jupiters could be produced by the
Kozai mechanism.

\subsection{Predictions of the Kozai Migration Scenario}
\label{subsec:predict}

The number of hot Jupiters produced by Kozai migration can be
determined by observations in the near future, since Kozai migrated
planets must have a number of attributes. First, candidate Kozai hot
Jupiters will reside in binary star systems, although the binary mass
ratio may well be small; a brown dwarf companion can be dynamically as
effective as a solar-type companion (eq. [\ref{eq:bcrit}]). The study
by \citet{Mayor} establishes that $\sim 60\%$ of the stars in the
solar neighbourhood are actually binary or triple systems.  While
radial velocity surveys select against close binaries, studies by
\citet{Raghavan} show that at least $23\%$ of radial velocity planet
hosts have stellar companions. The discoveries of brown-dwarf
companions to the planet bearing stars HD 3651 \citep{Mugrauer} and HD
89744 \citep{Mugrauer2} highlight the possibility that the existence
of dim companions will increase the known binary fraction of planet
bearing stars significantly. %
The Kozai scenario predicts that the
binary fraction of hot Jupiters will be higher than that of systems
with more distant planets.  
Binary-induced radial velocity trends induced on the primary by a
stellar companion will be of order
\be  
5f\left({m_c\over 0.3M_\odot}\right)
\left({100AU\over a_c}\right)^2\m/\s/yr,
\ee  
where $f$ is the sine of the angle between the line of sight and the
stellar velocity. This is clearly detectable at the current
sensitivity of radial velocity surveys \citep{Wright07}. The companion
will also induce an astrometric acceleration of a few
micro-arcsecond/yr/yr, detectable by SIM or GAIA.

%

Second, Kozai systems have 
$I \in [30^\circ,150^\circ]$, with $I\approx 90^\circ$ not uncommon
(Fig. \ref{fig:fig4}). In transiting systems the binary orbit will be
in or near the plane of the sky.  This can be tested via both radial
velocity and astrometry. %




Third, the angle between the spin axis of the primary star (assumed to
be the orbit normal of the planet at formation) and the present-day
planet orbit normal will range from 0 to 130 degrees
(Fig. \ref{fig:fig4}) with the values between 0 and 50 degrees being
preferred.  This angle can be determined if both the spin period of
the star as well as its rotational velocity $v\sin i$ can be
independently measured. The angle projected onto the plane of the sky,
measurable using the Rossiter-McLaughlin effect, will have a similar
range.

Fourth, the semimajor axis ratio $a_p/a_p'$ with any second planet
will be small. This results from the requirement that the precession
rate induced by the second planet not break the Kozai resonance
\citep{WuMurray}. Radial velocity measurements can constrain the mass
and semimajor axis of any nearby planetary companions to hot Jupiters
\citep{Wright07}. A corollary is that the fraction of multiplanet
systems having hot Jupiters will be smaller than the fraction of
single planet systems with hot Jupiters.


%
Kozai-migrated planets dissipate many times their own binding energy
during tidal circularization.  \citet{OgilvieLin} find that tidally
dissipated energy is deposited throughout the bulk of the planet,
raising the possibility that the planet will expand
catastrophically. In contrast, \citet{WuTide} concludes that energy is
deposited exclusively near the photosphere, which would leave the
planet intact.
The theoretical situation is unclear, but the existence of hot
Jupiters suggests an answer. A plot of $e_p$ versus $a_p$ strongly
suggests that the low $e_p$'s of the hot Jupiters are the result of
tidal circularization, 
{as the observed $e_p$'s follow closely the upper-bound set by the
tidal process \citep[see, e.g., Fig. 1 of][]{Wu03}.} If so, most or
all hot Jupiters have experienced rapid tidal heating and survived.
%

Another concern with the Kozai picture is raised by the
Rossiter-McLaughlin measurement of stellar obliquity, currently
available for 5 transiting planets \citep[see Table 2
of][]{Fabrycky}. All are consistent with zero obliquity.
Taken at face value, this is at variance with the above Kozai
prediction.
\footnote{HD147506 \citep{Winn,Loeillet}, a $1.3 M_\odot$
star with a massive planet, may have experienced tidal synchronization
in its surface-layer that would alter its apparent rotation axis.}

\subsection{Alternatives to Kozai Migration}
\label{subsec:alternative}

In Kozai migration, it is important that $r_p$ evolves on a time scale
no shorter than the tidal precession time scale; if $r_p$ were to
suddenly plunge from above to below the Roche radius, as for example
would be the case if two planets suffered a close encounter, the
inward scattered planet would not be stalled outside $r_R$. Instead
it would suffer rapid mass loss and likely be lost. In that case there
will be a cut-off in the distribution of $a_p$ at $2
r_R$\citep{FordRasio}, but not a pile-up.

Migration in a gas disk may also produce hot Jupiters.  If the disk
extends all the way to the star, one would observe a cut-off at $a_p
\sim r_R$; if the disk is truncated, e.g., by stellar magnetic fields
\citep{Linetal}, a feature will appear at an orbital period half that
of the inner edge of the disk.  However, spin periods and magnetic
fields of accreting stars show a substantial dispersion, which would
lead to a rather broad distribution in the disk inner radii, and hence
a smeared out feature in the distribution of planetary semimajor axis.

We have studied the role of a binary companion in increasing $e_p$ and
causing a gradual collapse in $r_p$.  But it is also plausible that
soft planet-planet scattering can gradually decrease $r_p$
\citep{JuricTremaine,Fordetal}. Moreover, Kozai oscillations can also
be excited by a second planet,\footnote{This second planet can be
placed on a highly inclined orbit by, e.g., planet-planet scattering.}
in the absence of a binary stellar companion. As long as these or
other processes produce gentle eccentricity driving on $10^4$ to
$10^8$ year timescales, tidal effects will halt the periapse evolution
when $r_p
\sim r_R$. Tidal circularization then pushes the planets out to $a_p
\sim 2 r_R$ and produces a narrow pile-up of hot Jupiters there.

\acknowledgements 
We acknowledge helpful discussions with Scott Gaudi, Daniel Fabrycky
and Andrew Gould, as well as NSERC discovery grants to YW and NM, and
an NSERC undergraduate fellowship to MR (summer 2006).


\end{document}